\documentclass[10pt,final,journal,twocolumn]{IEEEtran}
\ifCLASSINFOpdf
\else
\fi
\usepackage{cite}
\usepackage[cmex10]{amsmath}
\usepackage{amsthm}
\usepackage{algorithmic}
\usepackage{stfloats}
\usepackage{multicol,multienum}
\usepackage[dvips]{graphicx}
\graphicspath{{./figurespdf/}}
\usepackage{algorithm} %format of the algorithm
\usepackage{algorithmic} %format of the algorithm
\usepackage{multirow} %multirow for format of table
\usepackage{amsmath}
\usepackage{xcolor}
\usepackage{makecell}
\usepackage{subfigure}

\usepackage{amssymb}
\usepackage{color}

\begin{document}

\title{A Coalition-Based Communication Framework for Task-Driven Flying Ad-Hoc Networks}

\author{Dianxiong~Liu,
        Jin~Chen,
        Hong~Li,
        Yang~Yang,
        Lang~Ruan,
        Yuli~Zhang
        and~Yuhua~Xu

\thanks{Dianxiong Liu, Jin~Chen, Yang~Yang, Lang Ruan and Yuhua Xu are with the College of Communications Engineering, Army Engineering University, Nanjing, China. }
\thanks{Hong Li is Nanjing Military Representative Bureau of PLA Rocket Force, Nanjing, China.}
\thanks{Yuli Zhang is with National Innovation Institute of Defense Technology, Academy of Military Sciences PLA China, Beijing, China.}

%\thanks{Qihui Wu is with the College of Electronic and Information Engineering, Nanjing University of Aeronautics and Astronautics, Nanjing, China (e-mail: wuqihui2014@sina.com).}
%\thanks{Alagan Anpalagan is with the Department of Electrical and Computer Engineering, Ryerson University, Toronto, Canada (e-mail: alagan@ee.ryerson.ca).}
}

\IEEEpeerreviewmaketitle
\maketitle

\begin{abstract}
  In this paper, we develop a task-driven networking framework for Flying Ad-hoc Networks (FANETs), where a coalition-based model is outlined. Firstly, we present a brief survey to show the state-of-the-art studies on the intra-communication of unmanned aerial vehicle (UAV) swarms. The features and deficiencies of existing models are analyzed. To capture the task-driven requirement of the flying multi-agent system, a coalition-based framework is proposed. We discuss the composition, networking mode and the classification of data transmission. After that, the application scenario of UAV coalitions is given, where large-scale, distributed and highly dynamic characteristics greatly increase the difficulty of resource optimization for UAVs. To tackle the problem, we design an intelligence-based optimization architecture, which mainly includes the game model, machine learning and real-time decision. Under the guidance of game theories and machine learning, UAVs can make comprehensive decisions by combining the previous training results with their sensing, information interaction, and game strategies. Finally, a preliminary case and promising open issues of UAV coalitions are studied.
\end{abstract}

\section{Introduction}
The unmanned aerial vehicle (UAV) technology promotes the development of multi-dimensional wireless communication networks. It can be used to support ground wireless networks \cite{UAVcommunication}. For example, UAVs can serve as mobile base stations (BSs) in cellular networks or delivery data among ground communication devices. Moreover, cooperating with each other, UAVs can form swarms and perform some dangerous tasks, such as target detection, disaster management and surveillance in remote areas.

For intelligent UAV systems, the self-organizing communication among drones is very important for the task implementation \cite{FANETs}. However, the internal communication of UAV networks were not well developed in existing researches. More often, UAVs are controlled by people on the ground in real-time, where the positions and flight paths are presupposed before flight, so that the communication within the swarm is not needed. It is impractical that all of UAVs connect with ground controllers directly due to the constrained hardware. Particularly in long-distance missions, there may not be a powerful real-time air-ground wireless link, and only partial UAVs can communicate with ground controllers. Therefore, self-organizing communications are required among UAVs, and Flying Ad-hoc Networks (FANETs) were proposed and designed \cite{FANETs}.

Characteristics of FANETs including high dynamic, large-scale, heterogeneous and task-driven, are quite different from the wireless communication architecture on the ground \cite{dynamicframework}. Many articles compared FANETs with Mobile Ad-hoc Networks (MANETs) and Vehicular Ad-hoc Networks (VANETs) regarding to communication devices such as the mobility models, spatial dimension and computational power \cite{FANETs}. In this paper, we further discuss the essential difference of communications in FANETs versus ground Ad-hoc networks.

On the one hand, the driving force of the communication in FANETs is different from the ground wireless network. Communication requirements of ground wireless devices are always generated spontaneously by themselves. Users have their own transmission needs, which we call it as the bottom-up requirement of communications. By contrast, drones need to collaborate on one mission in task-driven networks. Their objectives of communications are for better completion of tasks, where we call it as the top-down communication requirement. On the other hand, performing remote tasks without the assistance of terrestrial infrastructures, UAVs need to make self-organizing strategies including data transmission, networking and task implementation by learning the surrounding environment independently, which will be more challenging. The development of artificial intelligence (AI) technology and the improvement of UAV capability will enable various networking and optimization becoming more flexible \cite{matchine_learning}. It is important and promising to study the intelligent communication architecture and related technologies of UAV networks.

\begin{figure*}
  \includegraphics[width=5in]{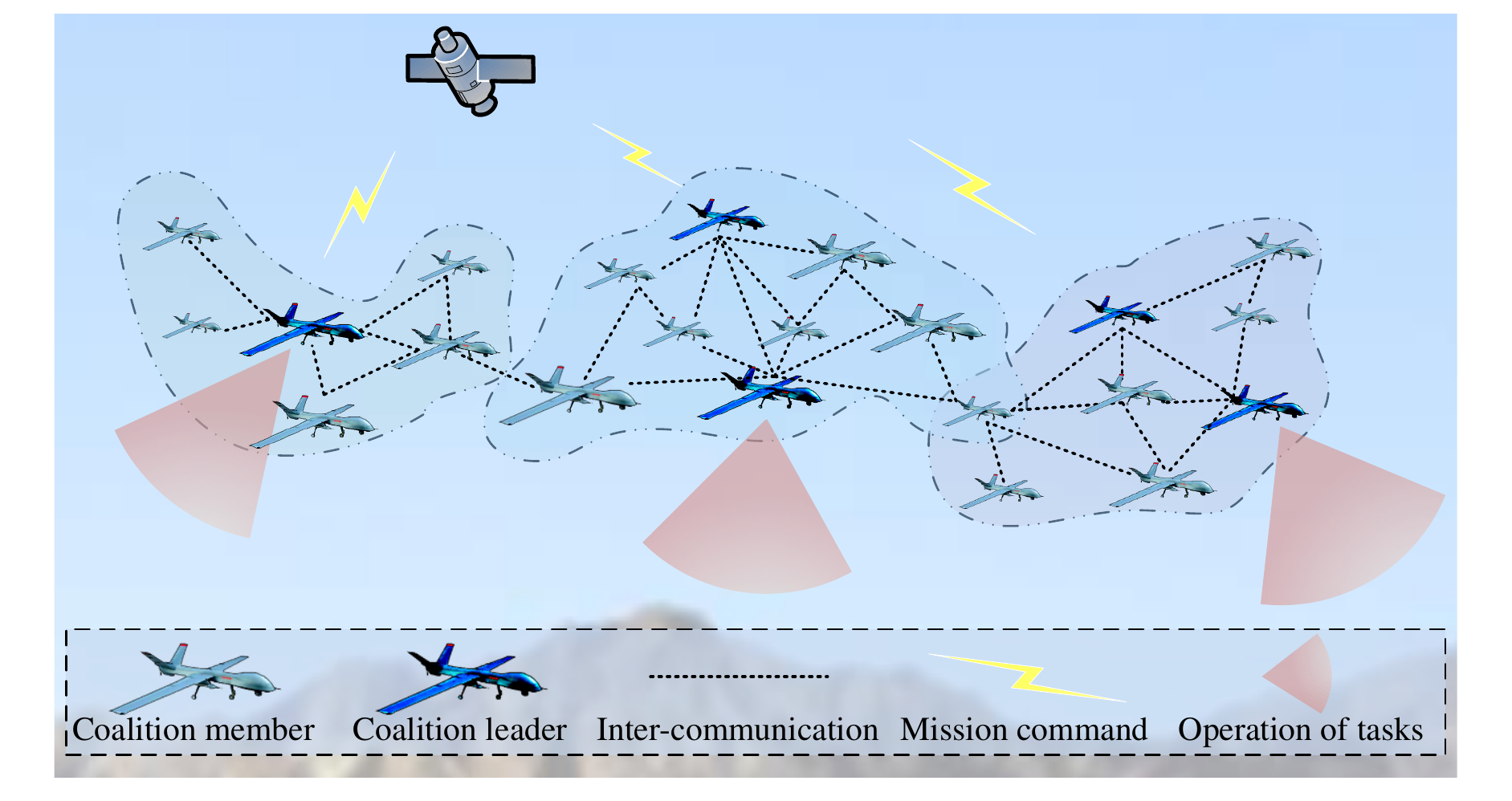}\\
  \centering
  \caption{The system model of coalition-based UAV communication networks.}\label{systemmodel}
\end{figure*}

We analyze FANETs from the perspective of intelligent optimization, where a coalition-based networking architecture is proposed for the internal communication of UAV networks. In the proposed model, UAVs are divided into several coalitions to take on various communication sub-tasks such as data collection and reconnaissance. Different from the previous architecture mainly focused on the quality of connection among UAVs, drones in the proposed coalition-based networks will be more autonomous to deal with the diversity of tasks.

In the following, we elaborate the framework of coalition-based UAV communications by answering three questions respectively: why the task-driven UAV communication networking is needed, what the coalition-based UAV communication framework is, and how to optimize the coalition-based model. The details and main contributions of this paper are as follows.

\begin{itemize}
  \item We give a brief review of the existing communication models in UAV swarms including pre-planned, centralized, purely distributed and clustering UAV communication networks, and then analyze their features, applicable scenarios and shortcomings, respectively. Aiming at the internal communication of intelligent UAV swarms, requirements of task-driven UAV networking are discussed. Based on this, a coalition-based UAV communication is designed.
  \item We discuss the proposed coalition-based model by introducing the composition (i.e., coalition leaders and coalition members), the pattern of organization, and communication modes. The proposed communication network is more able to cope with the requirement of FANETs, which is conducive to the implementation and optimization of tasks.
  \item We design an AI-based optimization framework which mainly consists of game models, machine learning and online decision mechanisms. Game theories provide a powerful theoretical direction for the learning and decision making \cite{game_learning,potential_game}. Machine learning, including deep reinforcement learning based on big data and transfer learning with sparse data, respectively, can be used to train optimization samples of UAV networks. Assisted by training models, UAV swarms are able to complete a series of optimizing tasks by sensing, information interaction as well as game-based strategies.
\end{itemize}

The rest of this paper is organized as follows. In Section II, after the discussion of existing multi-UAV networking frameworks, the requirement and challenge of intelligent UAV communications are analyzed. In Section III, the coalition-based communication model is presented to cope with task-driven UAV networks. In Section IV, for various application scenarios, the self-organizing optimization framework of UAV networks is designed. In Section V, the challenges and open issues are discussed. Finally, conclusions are given in Section VI.

\begin{table*}
\renewcommand{\arraystretch}{1}
\caption{The classification and characteristic of multi-UAV swarms.}
\label{table1}
\centering
\begin{tabular}{|p{3.5cm}|p{4cm}|p{3.8cm}|p{4.4cm}|}
\hline

\hline
\makecell[c]{Multiple UAVs networks} & \makecell[c]{Features} & \makecell[c]{Application~scenarios} & \makecell[c]{Shortcomings} \\
\hline
\makecell[c]{\multirow{3}[6]{*}{Preplaned UAVs \cite{Joint_Trajectory}}} & Without information exchange among drones, the pattern of networking is controlled by ground controllers. & Drone shows; small scale fixed models in stable environments (farms, playgrounds).  & Lack of emergency response capability and are vulnerable to destruction. \\
\hline
\makecell[c]{\multirow{3}[6]{*}{BS-connected UAVs \cite{dronesharing}}} & Three-dimensional wireless access equipment, where the resource of drones can be deployed by centralized methods. & Urban applications, such as advertising, hot spot covering and monitoring. & The additional burden on the BSs, where the large number of drones may overload the cellular network. \\
\hline
\makecell[c]{\multirow{4}{*}{Flat organizational UAVs \cite{Cooperative}}} & Pure distributed management and Self-organizing communication among drones, which belongs to FANETs. & Self-organizing networks, using in such as coverage deployment and position adjustment of multiple UAVs. & A purely distributed UAV network is difficult to adapt to large-scale scenarios in the complicated environment. \\
\hline
\makecell[c]{\multirow{2}[5]{*}{Cluster-based UAVs \cite{FANETs}}} & Semi-distributed network, in which UAVs are divided into several small communication networks. & Large-scale communication models to ensure the transmission of security information. & Driven by bottom-up demands of communications, cannot meet the requirement of task-driven UAV swarms. \\
\hline

\hline
\end{tabular}
\end{table*}

\section{Why the Coalition-Based Network Is Designed}
Multiple UAVs form as groups and coordinate with each other for correlative tasks. If UAVs belong to the same operator, they need to obey orders given by the operator. Even if they come from different operators, win-win situations among drones also need the inter-coordination in UAV swarms. We give a simple review of the existing multi-UAV swarms, and then compare the feature of intra-communications in different frameworks (as illustrated in Table I).

\subsection{The classification of existing communication models for multi-UAV swarms}
\begin{enumerate}
\item Preplanned UAV swarms: Without self-decision making, missions that UAVs carrying are preplanned by the ground controller before flying, including the trajectory, hovering position, and the coverage area \cite{Joint_Trajectory}. These kinds of UAV swarms are suitable for working in a stable flying environment, but not for modeling dynamic networks, and also lack emergency measures in actual operation.
\item BS-connected multiple UAVs: Existing researches mainly focused on ground-based UAV communications. Besides served as air BSs, some of researches assumed that UAVs and ground users share the spectrum resources of the cellular network \cite{dronesharing}. Such UAV networks rely on the ground infrastructures strongly. The increasing number of UAVs will cause over burden to ground wireless networks.
\item Flat organizational UAV networks: Without the reliable help of ground controllers, drones achieve the self-organizing operations by pure distributed optimizing methods \cite{Cooperative}. This type of issues occurs in such as the coverage deployment and position adjustment of multiple UAVs. Drones coordinate the spectrum access and tasks performing by sensing and interacting information with other drones.
\item Clustering UAV communications: The cluster-based framework is an important networking architecture of FANETs \cite{FANETs}, mainly used in networks with limited support from operators. In cluster-based models, cluster heads are responsible for connecting with ground controllers. They share the resource with other cluster heads in distributed ways. Within one cluster, however, resources of cluster members can be assigned by cluster heads so that centralized schemes can be used.

\end{enumerate}

Preplanned and centralized models do not involve the intra-communication of UAV swarms. By contrast, UAV swarms are self-organizing in FANETs, which is more suitable to operate remote tasks without strong control of the operator. It designs communication protocols of the physical layer, medium access control (MAC) layer, network layer and transport layer according to the characteristic of UAV swarms. The FANET is seen as a subclass of the VANET and MANET \cite{FANETs}, and is special in terms of mobility, message routing, topology change, application areas, etc. Similar with ground networks, the cluster-based model was proposed. However, VANETs and MANETs are modeled to guarantee individual needs, where nodes move and communicate spontaneously. FANETs are constructed to finish tasks, where UAVs are directed by one or relative multiple ground controllers. The clustering model in ground networks may not be suitable for UAV networks which are not only required to achieve safe flying, but also intelligent organization for operating missions.

\subsection{The shortage of clustering models vs. the requirement of task-driven UAV networks}
We describe shortcomings of the existing cluster-based framework in task-driven modeling, and discuss the demand of self-organizing UAV networks, as follows:

\begin{enumerate}
  \item In each cluster, there is at most one head responsible for the communication among clusters and the management within its cluster \cite{VANETcluster}. However, the cluster head connected with the ground controllers may not in a favorable position to guide task implementation. Driven by tasks, UAV mission groups can have more than one head, responsible for communication with ground controllers and the implementation of specific tasks, respectively.
  \item The clustering architecture is relatively simple in VANETs and MANETs, which divides users into independent communication groups without considering the task cooperation. If referring to the ground cluster-based network \cite{VANETcluster}, one node cannot serve more than two clusters at once. By comparison, UAVs are split into several teams according to various mission requirements, so they may be attached to one or more mission groups to complete various transmission tasks.
  \item Clustering between drones is generally based on the distance and communication quality. Cluster heads contend for the spectrum resource and assign them to members within the cluster, which is a kind of semi-centralized networking. For UAV communications, members leave or join UAV teams according to the task operating. An intelligent networking framework is required, where the formation and split of UAV groups can be more diverse and rapid.
\end{enumerate}

Therefore, the task-driven UAV swarm is different from VANETs and MANETs obviously, which can not be neglected. The cluster-based architecture cannot meet the requirement of task-driven communications among drones. In the next section, we present a coalition-based communication model, where drones are seen as multiple intelligent agents. The collaboration among intelligent agents is still an open issue of AI, and the coalition model is seen as an efficient model in multi-agent systems \cite{multi-agent}. Combining features of the coalition model with the requirement of UAV swarms, the proposed model is more adapted to the task-driven characteristics of UAVs, as detailed below.

\section{What Is The Coalition-Based Network}
\subsection{Elements of the coalition-based model}
As shown in Fig. \ref{systemmodel}, satellites are used to transfer the information between ground controllers and UAV networks. Driven by mission requirements, multiple UAVs are divided into several coalitions which consist of coalition leaders and coalition members.

Coalition leader: Coalitions should maintain the connection with ground controllers to receive task commands and feed back the real-time information. Because not all drones can be equipped with such heavy hardware modules \cite{FANETs}, there is a part of UAVs that can be selected as the group head to gather data and communicate with command centers. Moreover, during task performing such as detection and tracking, drones which are close to the target or easy to mobilize the coalition can take a lead role in the mission-aware communication. Task-based intelligent UAV swarms need coalition leaders to guide these two different types of communications. Therefore, as shown in Fig. \ref{systemmodel}, coalition leaders can be two independent drones or a single drone.

Coalition member: UAVs join or leave certain coalitions according to task requirements. Due to the complexity and correlation of tasks (such as detection and rescue missions), there may be overlapping between different coalitions \cite{overlapping_coalitions}. Thanks for the improvement of the device communication module such as multiple-input multiple-output (MIMO), drones can access various channels to participate in several communication coalitions at once, so that they can coordinate transmission and task scheduling according to different mission requirements.

\subsection{The driving force of the coalition networking}
The existing clustering model is generally based on the communication distance to ensure the safety information transmission between users \cite{VANETcluster}. In contrast, the formation and division of coalitions are more about needs of swarms and the task guidance. There are three cases of coalition networking, which are driven by command control guidance, task communication guidance and emergency communication, respectively.

\begin{figure}
  \includegraphics[width=3.5in]{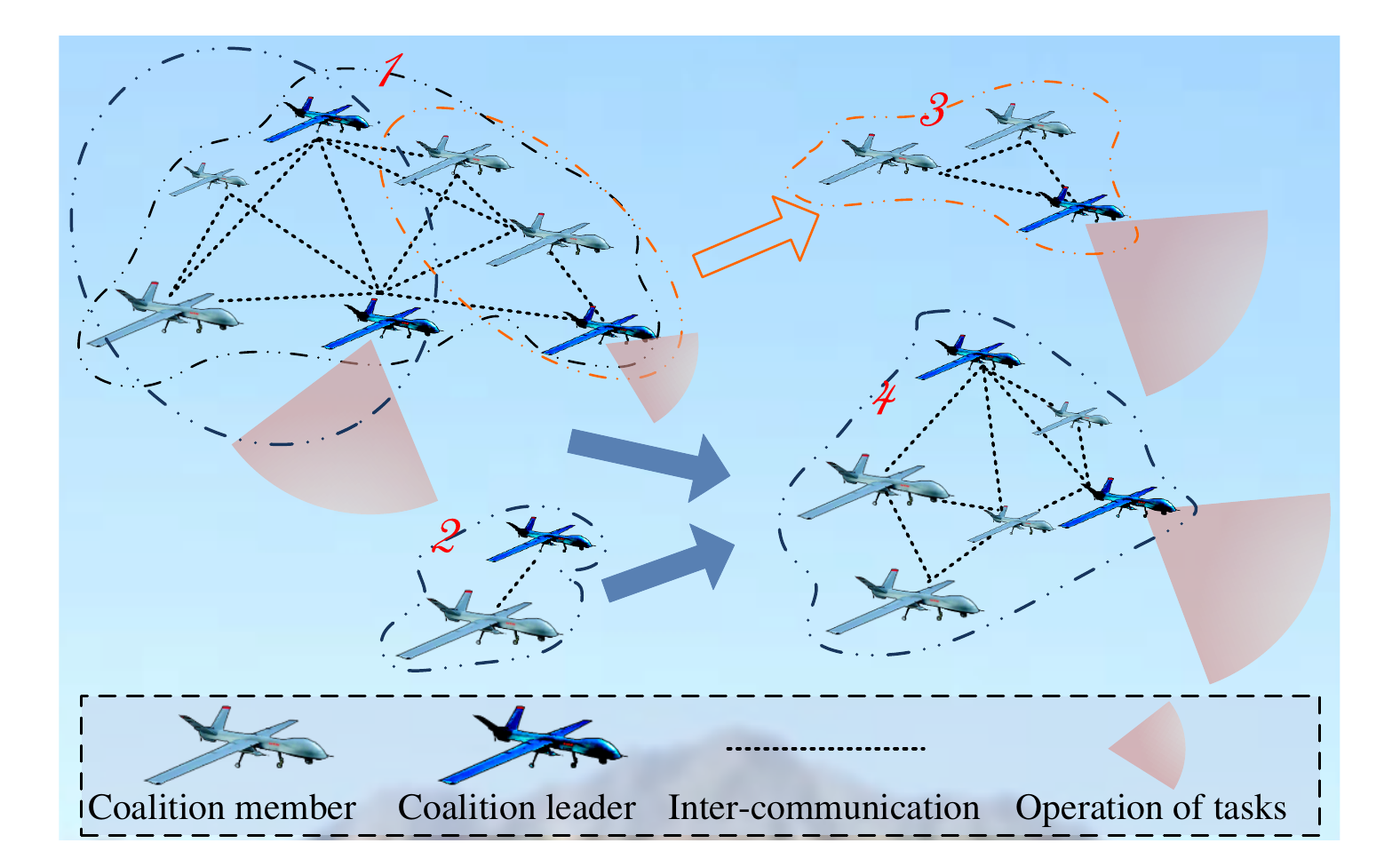}\\
  \centering
  \caption{An example showing the formation and split process of UAV coalitions.}\label{formationandsplit}
\end{figure}

Command control guidance: The ground command center will be able to replan missions of UAV coalitions after receiving the feedback information. For example, as shown in Fig. \ref{formationandsplit}, No. 1 coalition is reassigned two different missions, so a part of drones split from the original group and form as another coalition (No. 3 coalition in Fig. \ref{formationandsplit}). The ground controller directs the restructuring of coalitions to help the assignment of tasks and ensure reliable links between drones and command systems.

Task communication guidance: The convergence and separation of missions will lead to the merger and dissolution of communication coalitions. If there is a crossover or similarity between various tasks, task guiding leaders can make decisions to merge the coalition by information interaction. In Fig. \ref{formationandsplit}, the remainder No. 1 coalition and the No. 2 combine to form a new coalition. With the change of coalitions, coalition leaders need to be re-elected, and relative members also choose whether to join or leave the coalition.

Emergency communication: When swarms suffer from severe communication disturbance, UAVs will broadcast the security information to neighboring drones to ensure the safe distance and the flight path. They can also form temporary coalitions to guarantee the basic communication with ground controllers to keep the stable structure of UAV swarms.

\subsection{The classification of coalition-based UAV communications}
Information interaction is required among drones to coordinate tasks. We categorize coalition-based UAV communications according to information types and the transmission range.

\subsubsection{Safety information broadcast among adjacent UAVs}
Similar to VANETs, the information broadcast of location and trajectory within the group is needed to ensure the safe flight of multiple UAVs. The safety information exchange generally occurs within coalitions. In addition, between or belonging to two coalitions, drones can broadcast and listen to the location information with surrounding UAVs by multiple transceivers.

\subsubsection{Information fusion communication within coalitions}
The data of tasks are required to broadcast or transmit within coalitions. Task guiding leaders should assign missions to other drones. Ground connecting leaders need to summarize the information of UAVs within the coalition and then transmit to the ground controller. If coalition members are located far away from each other, the communication can be carried out in the relaying manner \cite{dynamicframework}.

\subsubsection{Shared and cooperative communication between coalitions}
Adjacent coalitions need to compete and coordinate spectrum resources according to the communication requirement. The information interaction between UAV coalitions helps the coordination of tasks, which is less time-consuming and flexible than that of satellite forwarding. Moreover, some coalitions do not have the ability to communicate with the ground controllers directly, so they require help from other coalitions to transfer data. Among them, the information can be transmitted through coalition leaders or members between different coalitions.

\begin{figure*}
  \includegraphics[width=5in]{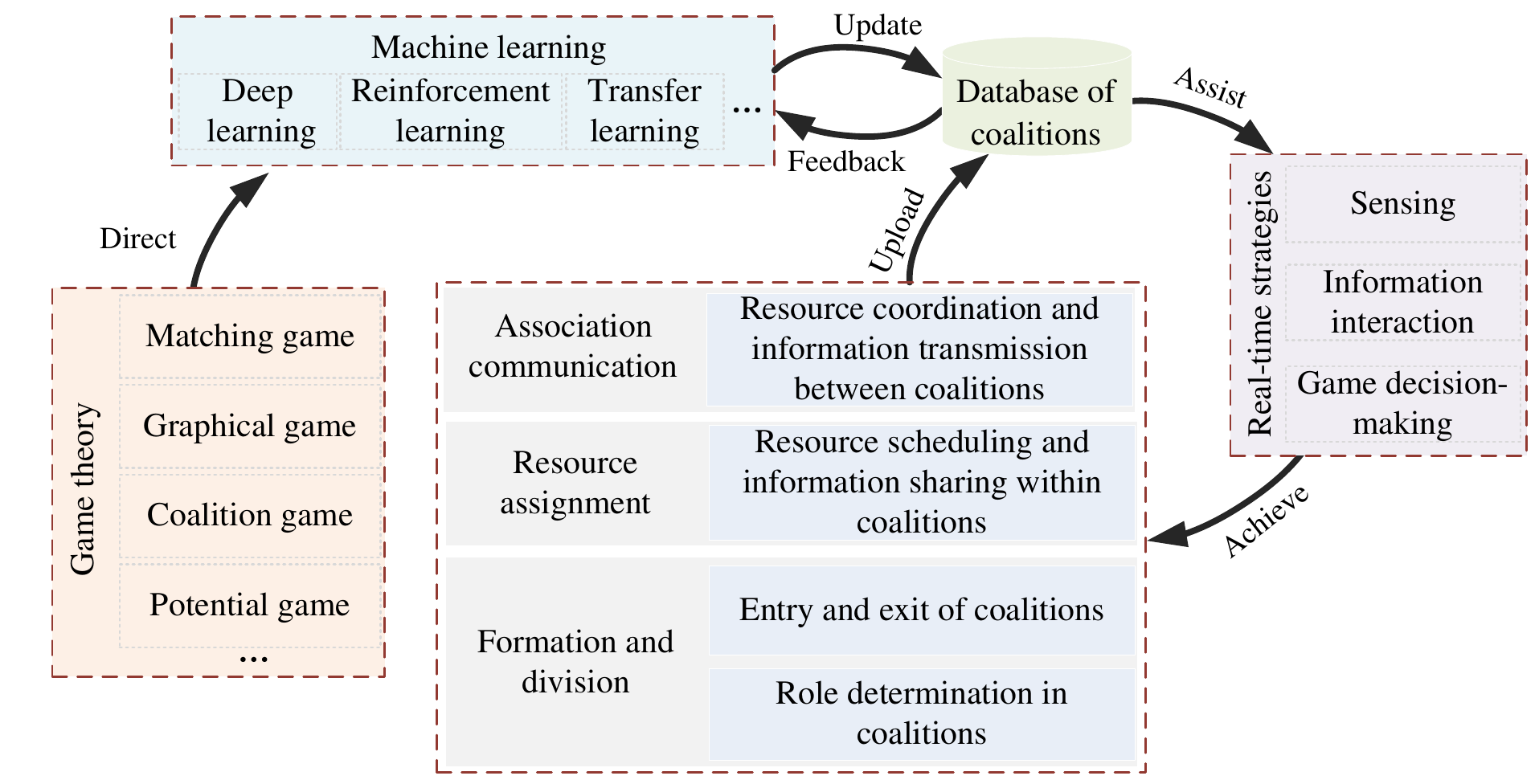}\\
  \centering
  \caption{The AI-based optimization framework including the theoretical guidance, machine learning and real-time strategies.}\label{framework}
\end{figure*}

\section{How to Optimize The Coalition-Based UAV Networks}
The application of UAV swarms has been discussed in existing researches, such as reconnaissance, search and rescue \cite{UAVcommunication}. Here, we give an example from the view of task implementation. During the disaster searching, if field situations are unknown by the ground controller, the aircraft group is supposed to collect and feed back the real-time information, even carry out the rescue work. UAVs that discover the target can assemble nearby drones to form a communication coalition. In the coalition, one drone who communicates well with the ground controller can be elected as the ground connecting leader. Coalition members can be affiliated with more than one coalition to ensure the completion of various tasks such as reconnaissance and rescue. Coalitions need to coordinate with each other to ensure the completion of missions in the shortest possible time.

\subsection{The optimization framework of UAV coalitions}
Based on the coalition-based model, we discuss the optimizing technologies of self-organizing UAV networks. It can be predicted that, UAVs will face with complex terrain, weather and targets. Factors such as the dynamic environment, large-scale networking and compound tasks cause great challenges to the optimization. To tackle the problem of data transmission and resource utilization effectively, we design an AI-based optimization framework mainly comprised of the theoretical guidance, machine learning and real-time strategies. As illustrated in Fig. \ref{framework}, relevant game theory models can be constructed for different objectives to provide the reliable guarantee of distributed decision-making \cite{game_learning}. The sample space of UAV optimizing models can be trained through the machine learning and stored in the database of coalitions to assist the real-time decision. In the dynamic environment, UAVs can make strategies by matching the learning results with their own sensing, information interaction and game decision making comprehensively.

The game theory has been proved to be a powerful guideline to model the user relationship in distributed wireless communication networks \cite{game_learning}. For self-organizing UAV networks, various game models can be used to model the distributed coalition-based architecture. For example, the coalition game provides guidance for the merge, split and resource allocation of coalitions \cite{overlapping_coalitions}. The matching game can be used to model the service information transmission between UAVs \cite{dynamicframework}. The graphical game and potential game help to complete resource coordination, interference management and task assignment modeling among distributed groups \cite{potential_game}. By reasonably designing the utility function of game models (such as the potential function in the potential game, the preference rule in the matching game, etc.), the distributed optimizing method can achieve a good overall network optimization result \cite{game_learning,potential_game}.

%In \cite{Access}, the problem of database-assisted spectrum access in dynamic networks was studied by the game-based learning algorithm. In the distributed network, potential game help communication devices to learn the spectrum states and make access strategies without information exchange. It is shown in Fig. \ref{gameresult} that the game-based stability result can achieve the extent of the Nash equilibrium solution \cite{Access}, which is close to the optimal one by designing the utility function.

\begin{figure}
  \includegraphics[width=3in]{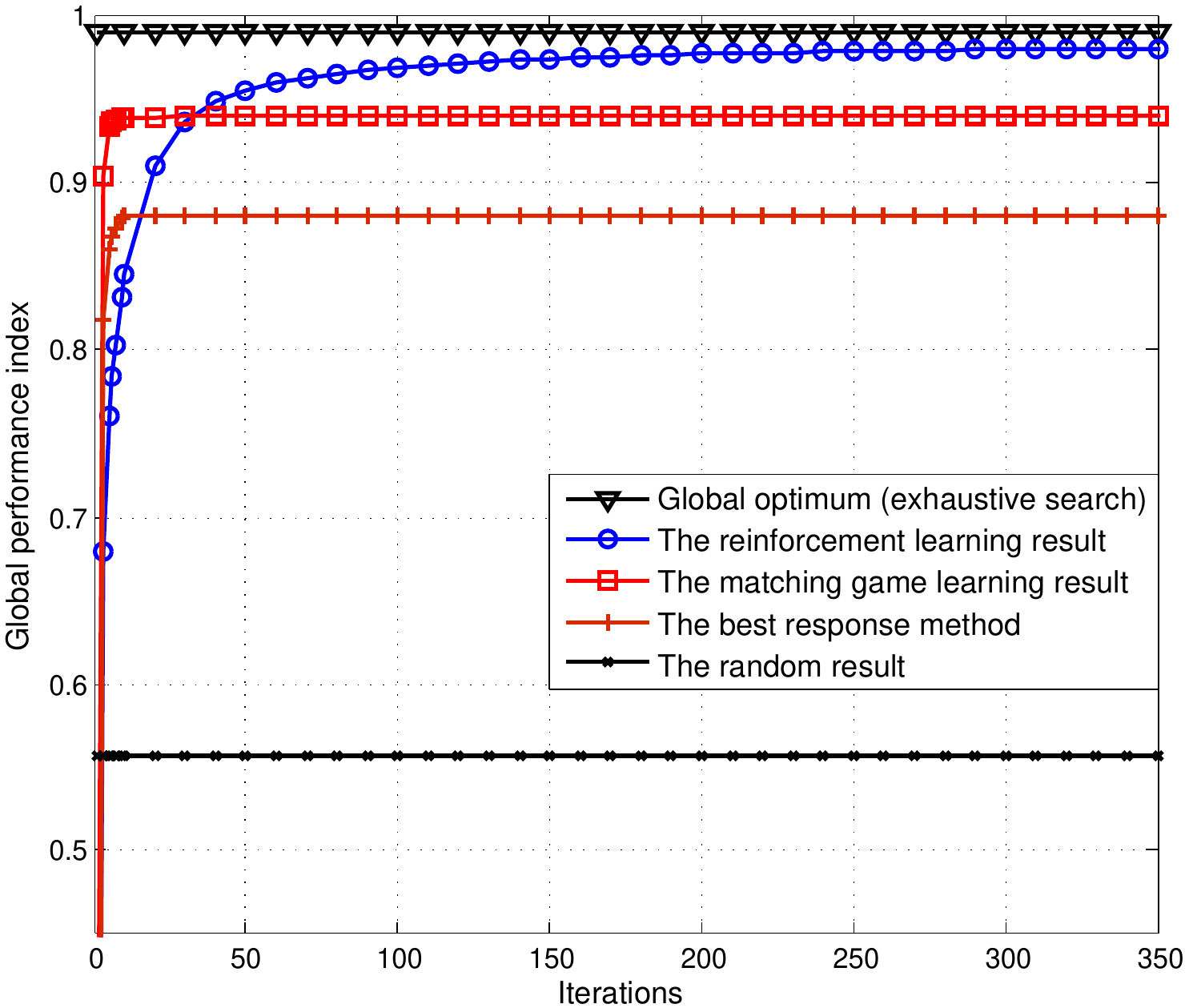}\\
  \centering
  \caption{Game-based learning algorithms applied in distributed UAV communication networks.}\label{varioslearning}
\end{figure}

Directed by the game model, the equilibrium solution, even the optimal result can be achieved through learning algorithms \cite{potential_game,manytomany}. In the previous work \cite{manytomany}, a matching game based model was constructed for drones to share the limited resource of relay drones. The learning algorithm under the guidance of the matching model can find the stable relay selection result by limited iterations. A simulation example is shown in Fig. \ref{varioslearning}, a near optimal result is achieved through the reinforcement learning with 350 time of iteration. At the same time, the suboptimal result is achieved by game learning algorithm with less than 20 iterations, which is rapid and can quickly respond to the dynamic communication environment.

It is seen that the machine learning is a powerful tool for implementing AI and has broad prospects \cite{matchine_learning}. Deep learning can be used to solve problems such as computer vision and speech recognition with the support of big data. For resource optimization problems, intelligent agents can use the reinforcement learning to find out the optimizing strategy in a certain network scenario after repeated iterations and attempts. With the help of deep learning, reinforcement learning is allowed to be applied in the large-scale optimization network (deep reinforcement learning) \cite{matchine_learning}. Besides, due to the complex and variable mission scenarios, it is very likely that some practical situations do not appear in existing data samples. Based on the transfer learning \cite{transfer}, knowledge learned from an optimizing sample can be used to help with task learning in the new environment.

Unlike ground cellular networks, applying deep reinforcement learning or other learning algorithms directly in distributed UAV networks may cost too much computation and iteration time, which is difficult for drones with weak computing ability. Machine learning can be generally carried out in the ground controller. Then the information of learned sample models and decision strategies can be stored in UAV coalition leaders. Coalitions leaders act as temporary databases to receive the command information from the ground controller on the one hand, and to facilitate the database call for coalition members on the other hand.

UAVs sense the surrounding environment and interact with neighboring drones to obtain decision information. With the help of the coalition database, UAVs make strategies, including spectrum utilization and coordinated transmission scheduling, by considering the current environment, task requirements and sample models comprehensively. Drones are required to feed back the decision information to the ground controllers through the ground connecting leader. For UAV networks in the stable environment, ground controllers can learn and correct current decisions so as to guide the adjustment strategies of drones. For dynamically unstable UAV networks, the feedback information can be used as a training sample which can assist in subsequent decision optimization.

%\subsection{Optimized evaluation index of task-driven UAV coalitions}
%The optimization objectives in ground communication networks are mainly throughput, transmission energy efficiency or the fairness of the network devices, which can also be used as standards to evaluate the transmission performance of UAV networks.
%
%Besides, because UAV swarms often have specific flight tasks in the air, we consider the completion of communication tasks required for flight missions as the evaluation criteria.
%Based on this, we propose Task completion index (TCI) which is a communication requirement from the top down mission than a communication requirement from the aircraft itself.

%\begin{figure}
%  % Requires \usepackage{graphicx}
%  \includegraphics[width=3in]{gameresult}\\
%  \centering
%  \caption{The effective optimization guidance for distributed communication networks by game models.}\label{gameresult}
%\end{figure}

\begin{figure}
  \includegraphics[width=3.5in]{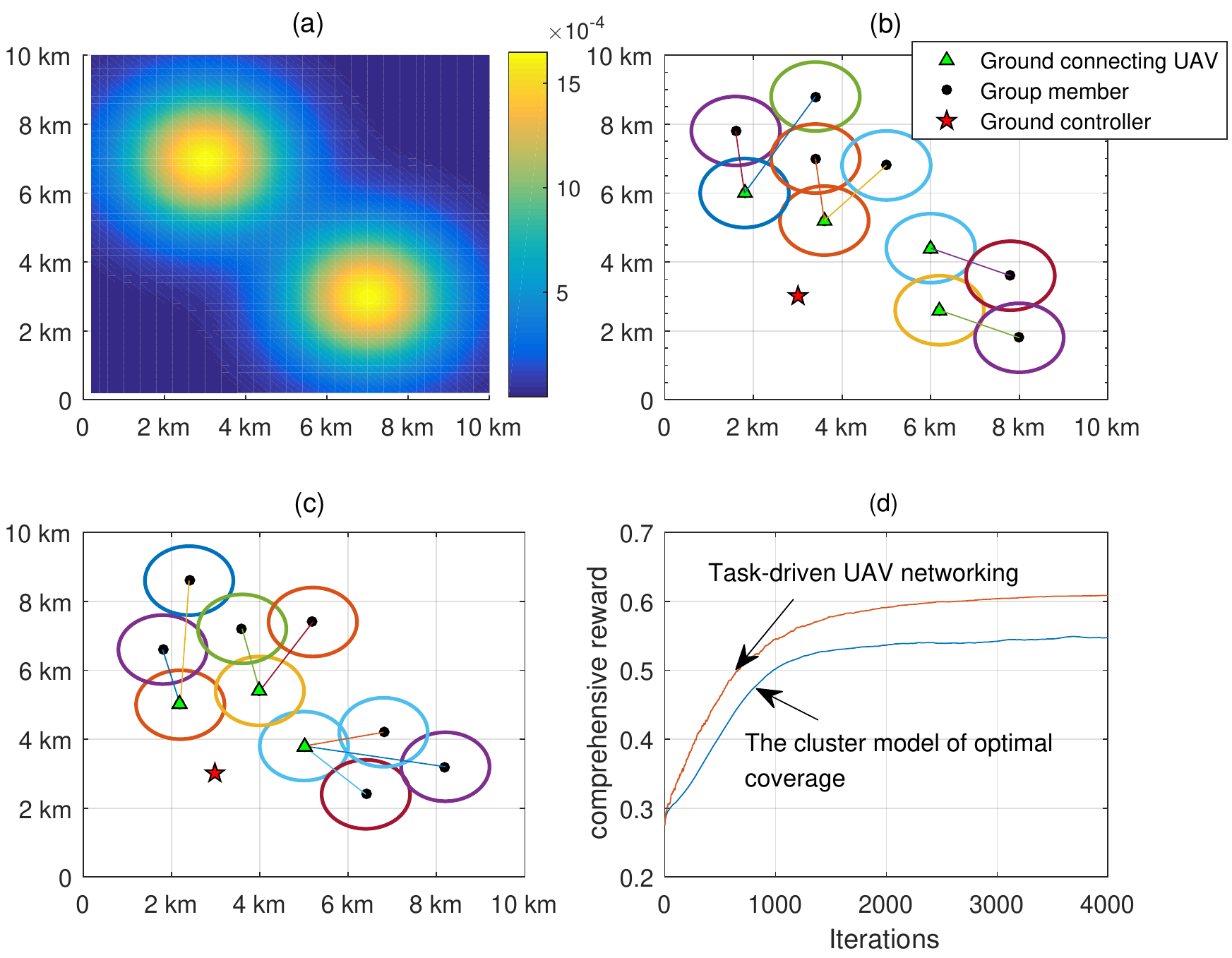}\\
  \centering
  \caption{A preliminary case study of the coalition formation among UAVs based on the game theory and machine learning.}\label{casestudy}
\end{figure}

\subsection{A preliminary simulation of UAV coalitions}

A preliminary case is studied \cite{previousworks} according to the above theoretical analysis. Assuming a fire accident occurs in a certain area ($10km\times10km$), as shown in Fig. \ref{casestudy}(a). When the ground controller is unable to access the accident, drones are required to perform missions of target detection and data collection. The search and rescue scope are extended centering on the place where the disaster occurred. As can be seen in Fig. \ref{casestudy}(a), the areas closer to the center position are more important, and the importance decreases gradually as the distance increases. With the constrained coverage ability, the UAV swarm needs to adjust the coverage area according to the importance of different areas, and it needs to collect the detection data back to the ground controller. Models only considering the performance of data transmission are difficult to achieve the requirement of coverage, while a network such as Fig. \ref{casestudy}(b) that only considers coverage performance results in a large transmission overhead. Therefore, driven by tasks, a multi-agent learning algorithm is designed based on the coalition formation game, where drones learn and adjust their location deployment, coalition formation and transmission overhead independently. By iteration learning, drones are divided into several coalitions shown as Fig. \ref{casestudy}(c) to complete tasks of the coverage and the transmission. Fig. \ref{casestudy}(d) shows that, the task-driven coalition model improves the optimization performance under the multi-index constraint.

\section{Challenges and Open Issues}
The proposed coalition-based model provides a new perspective to the resource optimization of task-driven UAVs, making the networking more flexible. Due to the heterogeneity and task complexity, more complex factors and characteristics will exist in the optimization of UAV communication networks:

\begin{enumerate}
  \item \emph{UAV coalitions considering heterogeneous characteristics}: In VANETs and MANETs, communication modules are similar between different wireless devices (cars and handheld devices, respectively). However, drones in UAV networks are very different in sizes, flying heights and communication modules. Some of them are allied for finishing tasks during flying, and some are appendages launched by large UAVs (such as ``Gremlins"\footnote{Available: http://www.darpa.mil/news-events/2015-08-28}). The modeling analysis should be performed in conjunction with the heterogeneity.
  \item \emph{UAV coalitions considering anti-jamming}: The stable operation architecture is important for UAV swarms in the interference environment, otherwise they may be captured by criminals, leading to a series of social and security problems. More and more intelligent interference modes put higher requirements on the anti-jamming optimization of distributed UAV networks. At present, there are preliminary studies using multi-user collaborative reinforcement learning and deep reinforcement learning to perform anti-interference training. This type of problem needs further research and utilization.
  \item \emph{UAV coalitions considering air-ground integrated networks}: The combination of UAVs and terrestrial wireless devices has been a concern for communications networks. The multi-dimensional extension of coalitions helps real-time decision making and flexible scheduling of tasks, and is also more resistant to destruction. For integrated networks, there are issues related to spectrum reuse and trajectory task scheduling, as well as monitoring mechanisms.
\end{enumerate}

\section{Conclusion}
This paper provided a novel networking framework for task-driven FANETs, where a coalition-based network was designed. Firstly, a brief review of the state-of-the art studies on the intra-communication of multi-UAV swarms was presented. It was shown that the requirement of task-driven UAV networks was not well studied in existing models. Therefore, this paper proposed the coalition-based communication framework for FANETs, and explained its features, including the composition, networking and data transmission. After that, the application scenario of the coalition-based network was introduced. Based on artificial intelligence, an optimization framework combining the game theory, machine learning and real-time decision making was designed. Finally, promising open issues of the proposed coalition-based UAV networks were discussed.

\end{document}